\documentclass[10pt,conference]{IEEEtran}

\usepackage{censor}
\usepackage[english]{babel}
\usepackage{colortbl}
\usepackage{comment}
\usepackage{graphicx}
\usepackage{epsfig}
\usepackage{array, colortbl} %
\usepackage{listings}
\usepackage{epstopdf}
\usepackage{multirow}
\usepackage{rotating}
\usepackage{float}
\usepackage[obeyspaces,hyphens,spaces]{url}
\usepackage{balance}
\usepackage{fancybox}
\usepackage{scalefnt}
\usepackage[normalem]{ulem}
\usepackage{cleveref}
\usepackage{xspace}
\usepackage{subcaption}
\usepackage{booktabs}

\usepackage[numbers]{natbib}
\makeatletter
\def\NAT@spacechar{~}%
\makeatother

\usepackage{nameref}
\usepackage{soul}
\usepackage{dirtytalk}

\newcommand{\lstbg}[3][0pt]{{\fboxsep#1\colorbox{#2}{\strut #3}}}
\lstdefinelanguage{diff}{
  basicstyle=\ttfamily\small,
  morecomment=[f][\lstbg{red!20}]-,
  morecomment=[f][\lstbg{green!20}]+,
  morecomment=[f][\textit]{@@},
}

\ifCLASSINFOpdf
\else
\fi

\usepackage{enumitem}
\setlist[enumerate]{leftmargin=*}

\definecolor{amber}{rgb}{1.0, 0.49, 0.0}
\definecolor{snsblue}{RGB}{86,125,185}

\newcommand\sbar[2]{{\color{snsblue}\rule{\dimexpr 1cm * #1 / #2}{6pt}}}

\newcommand{\npm}{npm\xspace}

\newcommand{\node}{\textit{Node.js}\xspace}

\newcommand{\stable}{not in decline\xspace}
\newcommand{\decline}{in decline\xspace}

\newcommand{\term}[1]{\textit{#1}}

\newcommand*\fullcaption[2][]{\caption[#1]{#1\xspace#2}}

\usepackage[
  disable,
  colorinlistoftodos
]{todonotes}
\usepackage{verbatim}

\makeatletter
\if@todonotes@disabled
\else
\paperwidth=\dimexpr \paperwidth + 6cm\relax
\oddsidemargin=\dimexpr\oddsidemargin + 3cm\relax
\evensidemargin=\dimexpr\evensidemargin + 3.5cm\relax
\marginparwidth=\dimexpr \marginparwidth + 3.5cm\relax
\fi
\makeatother

\newcommand{\diego}[2][]{\todo[#1,author=\textbf{Diego Elias},color=YellowGreen]{#2}}

\definecolor{ScarletRed}{rgb}{0.80,0.00,0.00}

\makeatletter
\newcommand{\labelname}[1]{%
  \def\@currentlabelname{#1}}%
\makeatother

\newcommand\mainpoint[2][.]{\vspace{.1in}\noindent\textbf{#2#1}}

\usepackage[skins]{tcolorbox}
\newcommand{\conclusion}[1]{\begin{center}\begin{tcolorbox}[skin=widget, left=2mm,right=2mm,top=2mm,bottom=2mm,boxrule=0.3mm,arc=0mm,coltitle=black,colframe=black!99!white,colback=white!88!gray,width=(\linewidth),before=\hfill,after=\hfill]#1\end{tcolorbox}\end{center}}

\usepackage{pifont}

\usepackage{tikz}

\makeatletter
\renewcommand{\paragraph}[1]{\noindent\textsf{#1}.}

\begin{document}

\title{Where to Go Now? Finding Alternatives for Declining Packages in the npm Ecosystem}

\author{
  \IEEEauthorblockN{Suhaib Mujahid}
	\IEEEauthorblockA{
		\textit{Mozilla Corporation}\\
		Montreal, Canada \\
		smujahid@mozilla.com}
	\and
	\IEEEauthorblockN{Diego Elias Costa}
	\IEEEauthorblockA{
    \textit{Université du Québec à Montréal}\\
		Montreal, Canada\\
		costa.diego@uqam.ca}
    \and
	\IEEEauthorblockN{Rabe Abdalkareem}
	\IEEEauthorblockA{
    \textit{Omar Al-Mukhtar University}\\
		Bayda, Libya \\
		rabe.abdalkareem@omu.edu.ly}
    \and
    \IEEEauthorblockN{Emad Shihab}
    \IEEEauthorblockA{
      \textit{Concordia University}\\
      Montreal, Canada \\
      emad.shihab@concordia.ca}
}

\IEEEtitleabstractindextext{
\begin{abstract}
    Software ecosystems (e.g., npm, PyPI) are the backbone of modern software developments. 
Developers add new packages to ecosystems every day to solve new problems or provide alternative solutions, causing obsolete packages to decline in their importance to the community.
Packages \decline are reused less over time and may become less frequently maintained.
Thus, developers usually migrate their dependencies to better alternatives.
Replacing packages \decline with better alternatives requires time and effort by developers to identify packages that need to be replaced, find the alternatives, asset migration benefits, and finally, perform the migration.

This paper proposes an approach that automatically identifies packages that need to be replaced and finds their alternatives supported with real-world examples of open source projects performing the suggested migrations.
At its core, our approach relies on the dependency migration patterns performed in the ecosystem to suggest migrations to other developers.
We evaluated our approach on the \npm ecosystem and found that 96\% of the suggested alternatives are accurate.
Furthermore, by surveying expert JavaScript developers, 67\% of them indicate that they will use our suggested alternative packages in their future projects.

\end{abstract}

  }

\maketitle

\IEEEdisplaynontitleabstractindextext

\IEEEpeerreviewmaketitle

\section{Introduction}
\label{sec:alternatives:introduction}
\IEEEPARstart{S}{oftware} ecosystems such as \npm, Maven, and PyPI save us from reinventing the wheel over and over by facilitating the reuse of code to implement common functionalities~\cite{Kikas_MSR2017}.
For example, the registry of the node package manager (\npm) alone hosts more than 2.3 million packages to date and is still growing exponentially~\cite{Latendresse_ASE2022}.
Leveraging packages from the ecosystem can boost development productivity~\cite{Abdalkareem_FSE2017}, and improve software quality~\cite{Zerouali_SANER2017}.
However, there is no such thing as a free lunch.
The large size and rapid increase in the number of packages in the ecosystem have drawbacks.
Developers need to spend time and effort to find the right package to use in their projects.
Also, since software, like people, gets old~\cite{Parnas_ICSE1994}, developers need to keep up with the changes in the ecosystem to avoid depending on packages that became obsolete, dormant, or even deprecated~\cite{Valiev_FSE2018}.

Community interest uphold packages to improve, i.e., include better features driven by community needs, keep up the package maintenance by reporting bugs to maintainers, motivate maintainers to continue supporting the package, and some times even financially support the maintainers on platforms such as GitHub Sponsors~\cite{GitHubSponsors:online} and Open Collective~\cite{OpenCollective:online}.
Packages that show a decline in community interest are usually used less over time, become less frequently maintained, and eventually, could become abandoned~\cite{Khondhu_OOS2013,Valiev_FSE2018}.
Moreover, a package's decline in community interest may indicate that a better solution is drawing attention in the ecosystem, and developers are migrating to a package that better suits their needs~\cite{Mujahid_TEM2021}.

Prior work examined projects that are unmaintained~\cite{Coelho_ESEM2018, Coelho_IST2020} and identified packages that lose popularity over time (i.e., are \decline)~\cite{Mujahid_TEM2021}.
Other studies proposed approaches for mining dependency migrations from software repositories to suggest alternatives~\cite{Teyton_WCRE2012,Teyton_SMR2014,Alrubaye_ICSE219,He_SANER2021}.
However, to the best of our knowledge, little attention has focused on suggesting alternatives to packages that are \decline, especially in the context of dynamic programming languages such as JavaScript.

Therefore, in this study we leverage the wisdom of the crowd in the software ecosystem to suggest alternatives to packages that are \decline.
Our approach uses dependency migrations from real-world projects to identify alternative packages.
Moreover, our approach suggests dependency migrations based on the community interest of the packages, by replacing the packages that are \decline with alternative packages that still maintain the community interest.

We evaluate our approach on the \npm ecosystem, the host of JavaScript reusable packages and the largest and most popular ecosystem to date~\cite{Decan_EMSE2019,StackOverflow_Survey:online}.
The popularity and scale of the \npm ecosystem make it an ideal candidate for our study.
We evaluate the accuracy and the usefulness of our approach in generating alternatives for packages \decline, through the following three research questions:
\begin{itemize}[leftmargin=0em]
	\item[]\textbf{RQ1:} \textbf{How accurate is our approach in suggesting alternative \npm packages?} (\Cref{sec:alternatives:accuracy})
       Our approach identified 152 dependency migration suggestions, of which 96\% are valid suggestions. 
       Suggestions include alternatives of 10+ different package categories, from packages that assist project builds to implementing user interfaces, showing the wide applicability of our approach. 

        \item[]\textbf{RQ2:} \textbf{When and why maintainers migrate to depend on the alternative \npm packages?} (\Cref{sec:alternatives:characteristics})
	    The majority of migrations (74\%) are primarily motivated to replace unmaintained dependencies, which could cause maintenance issues for software projects. Most migrations occur as a dedicated maintenance task (69\%), however, 31\% of migrations occur during other development tasks, i.e., fixing bugs (16\%), adding new features (8\%) and code refactoring (7\%).   

	\item[]\textbf{RQ3:} \textbf{How useful is our approach to JavaScript project maintainers?} (\Cref{sec:alternatives:usefulness})
	      We surveyed 52 JavaScript developers to assess the usefulness of our approach.
       We found that our approach provided new information about alternative packages for 54\% of the developers. On a 5-points Likert scale, developers recommend having a tool that utilizes our approach to suggest alternative packages with median~$=$~4. More importantly, 67\% of the developers confirmed that they would use our suggested alternative packages in their future projects.
	
\end{itemize}

\diego{We need to flash out the contributions a bit more.}
Our findings show that our approach is accurate and helpful to JavaScript developers.
The following are the key contributions of our paper:
\begin{itemize}
	\item Propose an approach to suggest alternatives for packages \decline using migration trends in the software ecosystem.  
	\item Empirically evaluate our approach accuracy on the \npm ecosystem, and investigate the characteristics of the dependency migrations suggested by our approach.
	\item Surveyed 52 expert JavaScript practitioners to assess the usefulness of our approach through the awareness of suggestions, perception of usefulness, and whether our tool would motivate action from practitioners.
	\item Support the replication and future research by making all of our datasets (i.e., collected data, analysis results, scripts) publicly available~\cite{suhaib_mujahid_2021_5548231}.
\end{itemize}

\section{Approach}
\label{sec:alternatives:approach}
\begin{figure*}[t]
    \centering
    \includegraphics[trim=29cm 12.2cm 29cm 12cm, clip,width=.65\linewidth]{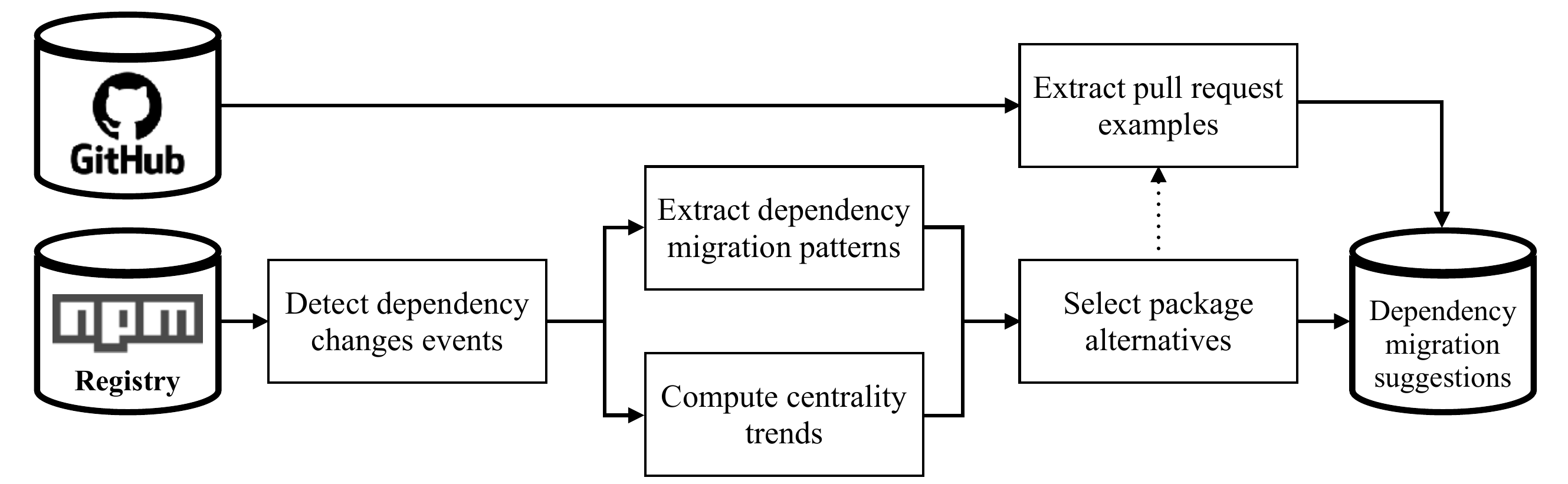}
    \caption{Our approach to suggest package alternatives.}
    \label{fig:approach_alternatives}
\end{figure*}

In this section, we explain our approach that uses the dependency migration patterns in the \npm ecosystem and the packages' centrality trends to suggest alternative packages for the ones that are \decline.
\Cref{fig:approach_alternatives} shows the overall workflow of our approach, which is further detailed in the remainder of this section.

\subsection{Detect Dependency Change Events}
\label{sub:detect_dependency_events}

The core of our approach lies in identifying frequent dependency migration patterns occurring in the ecosystem to better inform or recommend practitioners. 
To extract dependency migration patterns or compute the centrality trends, we need to detect \term{dependency change events} in the \npm ecosystem.
In our study, we consider two events as dependency change events: 1) the addition of a new package dependency and 2) the removal of a package dependency.
Hence, dependency change events do not include updating the version of a dependency, since our approach aims to suggest dependency migrations regardless of their versions.
Also, we use the dependency change events to update the \npm ecosystem's dependency graph and calculate the centrality trends (in \Cref{sub:calculate_centrality}).

To extract dependency change events for all packages in the \npm ecosystem, we analyze the entire \npm registry database.
The \npm registry maintains a record of the packages' dependencies for each version of every package.
For each package in the registry, we start by sorting the package versions in ascending order by their release time.
Then, on each version ($v_n$), we compare the list of dependencies with the previous version ($v_{n-1}$).
If the dependency is absent in the version $v_{n-1}$, we consider it to be a dependency addition event; conversely, if the dependency is absent in the version $v_n$, we consider it a dependency removal event.

It is crucial to note that package releases can be nonlinear.
Package maintainers commonly employ backports to fix older release versions~\cite{backport:21}.
In such cases, the chronological order of the versions will be polluted by backports, as these versions could include old dependencies no longer used in the main release branch.
Hence, in our process, we filter out any release with a lower semantic versioning than its predecessor in relation to their respective release date~\cite{Mujahid_TEM2021}.
For example, the developers of the package \texttt{react} have released the version \texttt{16.13.1} in March 2020, then in October 2020 released the backport version \texttt{15.7.0} to fix an older major version, where the latest version was \texttt{15.6.2}~\cite{react_npm:online}. Since the version \texttt{15.7.0} is smaller than the version \texttt{16.13.1}, we exclude the version \texttt{15.7.0} from our analysis.

\subsection{Extract Dependency Migration Patterns}
\label{sub:extract_migrations}

We extract dependency migration patterns by identifying recurring \term{dependency replacements} in the \npm ecosystems, where a dependency gets dropped in favour of an alternative package that performs the required functionalities.
First, we use the dependency change events to retrieve changes in the packages' dependencies across all their versions.
Then, we consider each of the added dependencies as a potential replacement for each of the removed dependencies.
\Cref{lst:dependencies_diff} show an example of dependency changes in a \texttt{package.json} file between two package versions.
The dependency changes in the example are represented as four dependency change events. Three dependencies are removed (i.e., removal change events), and one dependency is added (i.e., an addition change event).
By applying this on the example in \Cref{lst:dependencies_diff}, we extract three dependency replacements: $\texttt{less}\to\texttt{lodash}$, $\texttt{underscore}\to\texttt{lodash}$ and $\texttt{utf-8-validate}\to\texttt{lodash}$.
In our process, we do not mix runtime dependencies with development dependencies.
Thus, we consider the added runtime dependencies as potential replacements for only the removed runtime dependencies, not the development dependencies, and vice versa.
We do this since the development and runtime dependencies are typically used in different contexts and should not be recommended as alternatives to each other. 

\begin{lstlisting}[caption={Dependency changes taken from the version \texttt{0.2.16}  of the package \texttt{jpmorganchase/perspective}.},label={lst:dependencies_diff}, language=diff,float, frame=single]
    "dependencies": {
        "detectie": "1.0.0",
        "flatbuffers": "^1.10.2",
-       "less": "^2.7.2",
+       "lodash": "^4.17.4",
        "moment": "^2.19.1",
        "tslib": "^1.9.3",
-       "underscore": "^1.8.3",
-       "utf-8-validate": "~4.0.0",
        "websocket-heartbeat-js": "^1.0.7",
        "ws": "^6.1.2"
    },
\end{lstlisting}

When a package releases a new version and replaces more than one dependency, our approach has no way to identify which dependency has been replaced nor its replacement. 
Thus, as explained earlier, we consider each added dependency a potential replacement for each deleted dependency~(combination).
Since we want to reduce the odds of combinatorial explosion of dependency replacements, we filter out releases with massive or imbalanced number of dependency changes.
We only consider dependency change events from releases where the difference between the number of added dependencies ($D_a$) and removed dependencies ($D_r$) are close, i.e., $|D_a - D_r|\leq1$.
Also, we avoid considering change events where there is a large number of added and removed dependencies, i.e., $D_a+D_r\leq\tilde{\mathsf{\textit{x}}}$, where $\tilde{\mathsf{\textit{x}}}$ is the median value for $D_a+D_r$ across all the releases in the \npm registry.
We do this filtering since a large number of dependency changes indicates a significant code refactoring more than a simple dependency replacement.

We consider a \term{dependency migration pattern}, a dependency replacement that frequently occurs in the ecosystem, which is more likely to indicate a trend.
Hence, after identifying the dependency replacements, we consider replacements that reoccur at least 10 times in our dataset as dependency migration patterns. 
That is, at least the developers of 10 distinct projects must have performed the same dependency replacement.

\subsection{Calculate Centrality Trends}
\label{sub:calculate_centrality}

Centrality has been used as a proxy of community interest, where packages that show a decline in centrality are usually used less over time, become less frequently maintained, and eventually could become abandoned~\cite{Mujahid_TEM2021}.
Thus, our approach uses the centrality (i.e., measured using PageRank~\cite{Brin_PageRank1998}) to target suggesting alternatives for packages that are \decline to replace them with packages deemed \stable.
To determine if a package is \decline or not, we use the approach proposed by \citet{Mujahid_TEM2021} which requires dependency change events to calculate the centrality and detect packages \decline.
A package considered \decline if it shows statistically significant declines in the centrality over a specific period of time.

Our approach requires the centrality trends for packages engaged in the extracted dependency migration patterns.
However, calculating the centrality rankings requires computing the centrality for every package in the ecosystem.
Thus, we use the dependency change events (extracted in \Cref{sub:detect_dependency_events}) to calculate the monthly centrality rankings for each package in the \npm registry.

\subsection{Select Package Alternatives}
\label{sub:criteria}

Once we have both the dependency migration patterns (\Cref{sub:extract_migrations}) and centrality trends for all packages (\Cref{sub:calculate_centrality}), we select the most promising dependency migration patterns to recommend to practitioners.
To do so, we use the following criteria: 

\begin{itemize}
    \item \textbf{The replaced package is \decline:}
          we select only patterns where the removed package is \decline.
          Since the decline can vary based on the examined period, we measure the decline over three different periods: the last six months, the last year, and the package's overall lifetime.
          If the package shows a decline based on one of the measured periods, we consider it an \decline package.
    \item \textbf{The alternative package is \stable:}
            to ensure that we suggest better alternatives, we select only patterns where the added package is \stable.

    \item \textbf{The migration pattern is performed recently:}
          to avoid recommending outdated migration patterns, we consider only the patterns performed at least once in the last 90 days.
    \item \textbf{Performed by a popular project:}
          to avoid considering patterns performed only by immature projects, a migration pattern should be performed by a popular project in order to be considered.
          In this context, we consider a project as popular if the project is in the top 10\% of most central packages in the \npm ecosystem.
\end{itemize}

When our approach finds more than one package alternative, we select the dependency migration pattern with the highest support, i.e., performed more frequently.

\subsection{Extract Pull Request Examples}
\label{subsec:extract_pull_request_examples}
We aim to provide developers with examples of pull requests that performed the suggested dependency migration. 
Exemplary pull requests may provide insights on the migration's efforts, reasons for the dependency migration, and help practitioners understand the differences between the alternatives.
To extract pull requests examples, we check the \npm registry to find packages that performed any of the candidate dependency migration patterns.
For these packages, we collect their repository addresses on GitHub.
Then, we use the GitHub GraphQL API~\cite{GraphQL} to extract all the merged pull requests from the selected repositories.

Once we retrieve the pull requests from a repository using the GitHub API, we select only the pull requests that perform the suggested dependency migrations.
To do so, we consider only the pull requests that modify a \texttt{package.json} file, which is the file where projects declare their dependencies.
Then, we compare the content of the \texttt{package.json} file as it is on the merge commit with the content of the files as it was on the parent commit.
Next, we exclude pull requests that are extremely big, i.e., change more than 100 files, which is 7 times more that average number (mean $=$ 13.62) of changed files in pull requests~\cite{Gousios_MSR2014}.

\section{Accuracy of the Approach}
\label{sec:alternatives:accuracy}

The decline of package centrality is a symptom that better alternatives have emerged, shifting the community interest.
However, developers have little information to grasp where the community has shifted its interest~\cite{Mujahid_TEM2021}.
If our approach can effectively capture valid package alternatives, it can be embedded in dependency management tools, such as the \npm CLI, to increase developers' awareness of alternative packages and help them reevaluate their package dependencies.

\subsection{Approach.}
To measure the accuracy of our approach, we first use it to generate dependency migration suggestions to alternative packages.
Then, we manually evaluate whether the suggested alternative packages perform comparable functionalities to the original ones.

\noindent
\textbf{Generate Dependency Migration Suggestions.}
We generate the suggestions to alternative packages using the approach described in \Cref{sec:alternatives:approach}.
We start by detecting the dependency change events as of December 22, 2020.
We collected in total 18,459,923 dependency change events from 1,148,720 packages from the \npm ecosystem.
From these change events, we extracted 2,434 dependency migration patterns.
After filtering the migration patterns based on the centrality trend of the packages and the criteria described in \Cref{sub:extract_migrations}, we end up with 152~package alternative suggestions.

\noindent
\textbf{Manual Evaluation.}
Once we generate the dependency migration suggestions, we manually evaluate the correctness of the suggested alternative packages by assessing whether both packages provide similar functionalities or not.
We manually examine the documentation (e.g., readme file, homepage, and website) of the package to be replaced and the suggested alternative package to understand their functionalities.
We start by inspecting the documentation of the package homepage on the official \npm website~\cite{npmjs}.
If the description is not descriptive enough for our classification, we examine other available sources such as the readme file, the package website, and the package repository.
Once we have a comprehensive understanding of both packages, if the suggested alternative package performs comparable functionalities to the original package, we consider the suggested alternative package as a valid alternative package.
For example, the packages \texttt{commander} and \texttt{yargs} share similar functionalities, helping in building command-line interfaces for \node, thus, we consider them as valid alternative packages. 

In total, two of the authors independently examined the documentations of 256~packages for 152~dependency migration suggestions. Since this process involves human judgment, it is prone to human bias. We assess the agreement of both examiners using the Cohen-Kappa inter-rater reliability. 
Cohen-kappa inter-rater reliability is a well-known statistical method that evaluates the inter-rater reliability agreement level. The result is a scale that ranges between -1.0 and 1.0, where a negative value means poorer than chance agreement, zero indicates exactly chance agreement, and a positive value indicates better than chance agreement~\cite{Mary_HDMB2012}.
In our analysis, we found that both authors have an excellent agreement (kappa={0.79}).
After the initial classification, any disagreement was examined and discussed by both examiners to reach consensus~\cite{Fleiss_EPM1973}.

\subsection{Results}

\begin{table}[t]
    \centering
    \caption{Summary of the suggested alternatives categories.}
    \label{tab:suggestion_categories}
    \begin{tabular}{ll|rc}
    \toprule
    \textbf{Category} & \textbf{Suggestion Example}                    & \textbf{TP}  & \textbf{FP} \\
    \midrule
    Building          & $\texttt{browserify}\to\texttt{webpack}$ & 39           & 3           \\
    Utilities         & $\texttt{moment}\to\texttt{dayjs}$       & 26           & -           \\
    Testing           & $\texttt{vows}\to\texttt{mocha}$         & 16           & -           \\
    User Interface    & $\texttt{jade}\to\texttt{pug}$           & 12           & -           \\
    Linter            & $\texttt{jshint}\to\texttt{eslint}$      & 10           & 2           \\
    Client Library    & $\texttt{redis}\to\texttt{ioredis}$      & 10           & 1           \\
    Networking        & $\texttt{request}\to\texttt{axios}$      & 10           & -           \\
    Parser            & $\texttt{esprima}\to\texttt{acorn}$      & 9            & -           \\
    CLI               & $\texttt{optimist}\to\texttt{yargs}$     & 4            & -           \\
    Other             & $\texttt{memory-fs}\to\texttt{memfs}$    & 10           & -           \\
    \midrule
    \multicolumn{2}{l}{\textbf{Total}}                                                    & \textbf{146} & \textbf{6}  \\
    \bottomrule
\end{tabular}

\end{table}

Based on the manual evaluation of our approach, \Cref{tab:suggestion_categories} shows a summary of the migration suggestions generated by our approach.
We categorize and group the suggestions by their abstracted functionalities.
Each category of suggestions in the table has an example of a package migration suggestion generated by our approach.
The third column presents the number of True Positive cases (TP), where our approach accurately finds the alternative packages.
The last column shows the number of False Positive cases (FP), where our approach suggested invalided alternatives.

Overall, we examine 152 dependency migration suggestions generated by our approach.
We found that 146 (96\%) of the generated dependency migration suggestions include valid alternative packages and 6 (4\%) of them do not.
Furthermore, we found that the packages performing functionality related to building the JavaScript projects have the highest share (28\%) of the generated dependency migration suggestions.
For example, our approach suggests replacing the JavaScript bundler package \texttt{browserify} with a more popular, scalable, and feature rich one, the \texttt{webpack} package.
The next category in our results is the utility tools (17\%), where our approach suggests replacing obsolete utility packages such as \texttt{moment} with a more modern solution like \texttt{dayjs}.
Next categories include testing tools, and user interface components and helpers with share of 11\% and 8\% respectively.
Also, among others, the categories include linters to enforce rules on the JavaScript code, clients drivers to interact with other services, and networking utilities.

Out of the 6 invalid dependency migration suggestions that we found, only one invalid suggestion belongs to replacing a runtime dependency, whereas the remaining 5 cases suggest replacing development dependencies.
An example of invalid dependency migration suggestion is the suggestion of replacing the development dependency  \texttt{ember-cli-eslint} with \texttt{eslint}.
However, \texttt{ember-cli-eslint} is a plugin to identify and report patterns found in Ember projects~\cite{EmberjsA31:online}, not a valid alternative package of \texttt{eslint}.

\conclusion{
    \textbf{Summary of RQ1:}
    Out of the 152 dependency migration suggestions generated by our approach, we found that 96\% of them are valid alternative package suggestions.
    Most frequent suggestions recommend alternatives for building, utilities and testing packages. 
}

\section{Characteristics of the Suggestions}
\label{sec:alternatives:characteristics}

We want to understand \textit{why} and \textit{when} developers do the kind of package dependency migrations that our approach suggests.
Answering the question \textit{why} will help us find the reasons developers migrate to the alternative packages and how their motivation is tied to software maintenance. 
Answering the question \textit{when} aims to discover the types of maintenance activities involved in migrating dependencies. 
Both questions help us understand the reasons to use suggested migrations and how to employ our approach during software development.

\subsection{Approach}

To understand when and why developers migrate their dependencies, we manually classify pull requests that perform dependency migrations that match the migration suggestions by our approach.
We start by extracting pull requests using the approach described in \Cref{subsec:extract_pull_request_examples}. As a result, we obtain a list of 225 pull requests that perform dependency migrations from 155 different GitHub repositories.

Once we have the dependency migration pull requests, we perform an iterative coding process to classify and group pull requests.
We gradually develop two sets of codes based on an inductive analysis approach~\cite{seaman1999qualitative}.
The first set of codes concerned the purpose of the pull request (activity) and the second set of codes concerned the motivation of the dependency migration performed in the pull request.

In the process of classifying the activity of a pull request, we focus on the primary goal of the pull requests, as described in its title and description. Thus, we tag each pull request with only one activity type.
Also, we examine the pull request description and discussion to find the motivation of the dependency migration. However, not all pull requests include an explanation to justify the dependency migration. For the ones that have a justification, we tag them with the migration motivation.
Two authors independently tag each of the 225 pull requests with a single activity type, and applicable pull requests are also tagged with the migration motivation.

As with any other manual classification activity, there is some level of subjectivity that may generate disagreement between the annotators. 
To account for this, 
we applied a Cohen's Kappa to measure the level of agreement between the two individual classifications \cite{Cohen1960coefficient}. 
In our analysis, we found that both authors have an excellent agreement (kappa={0.93}) on classifying the pull request activates. Also, the authors have an excellent agreement (kappa={0.90}) on classifying the motivation of the dependency migrations.

\subsection{Results}

Based on the manual classification of the pull requests, we organize the results into two parts.
The first part discusses the motivation for migrating dependencies, and the second part dives into the types of maintenance activities performed during migration.

\begin{table}
    \centering
    \caption{The motivations of 62 pull requests that performed the dependency migrations.}
    \label{tab:pr_motivation}
    \begin{tabular}{lp{1.8in}|rl}
    \toprule
    \textbf{Motivation} & \textbf{Description}  & \multicolumn{2}{c}{\textbf{Frequency}}                 \\ \midrule
    Maintenance               & Better quality and better maintained alternative. & 74\%  & \sbar{45}{62} \\
    Compatibility             & Increase compatibility with other packages or systems. & 15\%  & \sbar{9}{62}  \\
    Performance               & Faster execution, less dependencies, or smaller bundle size.        & 8\%  & \sbar{5}{62}  \\
    Features                  & Providing missed features or flexible API.      & 3\%  & \sbar{2}{62}  \\
    \bottomrule
\end{tabular}

\end{table}

\begin{table}
    \centering
    \caption{The activities of the pull requests that performed the dependency migrations.}
    \label{tab:pr_activities}
    \begin{tabular}{p{.6in}p{1.8in}|rl}
    \toprule
    \textbf{Activity}   & \textbf{Description}    &                            \multicolumn{2}{c}{\textbf{Frequency}}                  \\ 
    \midrule
    Dependency update   & Intended mainly to update dependency versions.      & 42\% & \sbar{94}{225} \\
    Dedicated migration  & The main goal is to migrate to a different package. & 27\% & \sbar{61}{225} \\
    Bug fixing          & Aims to resolve issues in the project.                & 16\% & \sbar{35}{225} \\
    New feature         & Adds a new functionality or feature to the project. & 8\% & \sbar{19}{225} \\
    Refactoring         & The objective is refactor existing code.            & 7\%                                   & \sbar{16}{225} \\ 
    \bottomrule
\end{tabular}

\end{table}

\mainpoint[?]{Why developers migrate the dependencies}
Our manual assessment found that 62 (24\%) of the 225 pull requests provide an explicit justification of the dependency migration.
In \Cref{tab:pr_motivation}, we show the results of our manual classification.

In 74\% of the cases, dependency migrations are motivated by the need for better maintained alternative packages.
Even when the package is not officially deprecated, developers migrate from packages that are not well maintained, for example~\cite{Silbermann_PR_Maintenance:online}:
\say{Removes isomorphic-fetch from the dependencies which doesn't seem to be maintained anymore.}
Even when a package has maintenance activities, developers were not satisfied with the quality level of the maintenance, one of the developers said~\cite{Hardcastle_PR_Maintenance:online}: \say{Sentry is moving away from the Raven library we are using and while it says it's maintained, not every feature is working anymore.}

The second most frequent motivation for dependency migrations in our dataset is the compatibility with other packages or systems (15\%). In these cases, developers migrate to use alternative packages that are more compatible with other project dependencies or to support more systems and platforms. For example, in one of the pull requests the th developer migrated from the package \texttt{uglifyjs} to the package \texttt{terser} to be compatible with the new version of \texttt{webpack} package~\cite{Parsa_PR_Compatibility:online}:
\say{Webpack requires uglifyjs-webpack-plugin@1.x. thus uglifyjs-webpack-plugin@2.x may not resolve correctly. Also, the webpack team decided to go with terser-webpack-plugin.}
In another case, the migration performed to improve the support for the Windows operation system~\cite{Pakers_PR_Feature:online}:
\say{It is a nightmare to run this project on Windows as it uses bcrypt.}

We observe that 8\% of the dependency migrations are motivated by improving the performance of the project.
Performance metrics mentioned include faster execution time, smaller bundle size shipped to production, and lower number of transitive dependencies.
One developers improved the performance by migrating from the package \texttt{moment} to the package \texttt{dayjs}~\cite{Waterloo_PR_Performance:online}:
\say{\texttt{dayjs}, after webpacking, is about 7 KB compared to about 700 KB for \texttt{moment}. This will mean faster load times and smaller packed VSIXs.}

In the remaining of the pull requests (3\%), we observe that the motivation is to use features offered by the alternative packages. For example, a developer migrated from \texttt{sanitize-html} to \texttt{dompurify} to allow for more HTML tags after sanitizing HTML content~\cite{Slagle_PR_Fature:online}:
\say{\texttt{dompurify} prevents XSS but allows more tags and attributes than our previous sanitizer.}

\mainpoint[?]{When developers perform dependency migrations}
\Cref{tab:pr_activities} shows our classification results of the pull request activities.
We found that 42\% of the pull requests perform dependency migrations as part of a dependency update.
For example, in one of the projects, the developer created a pull request to update multiple dependencies, and in the same time migrated from using the package \texttt{node-uuid} to \texttt{uuid}~\cite{Bernhardt_PR_Update:online}.
Interestingly, we found that only 27\% of the pull requests are dedicated to performing a dependency migration activity.
In a dedicated migration, the main goal of the pull request is to just replace a dependency with another.
In contrast, a pull request tagged with dependency update activity aims to update the version of one or more dependencies, but it replaces other dependencies in the same pull request.

From our analysis, we identified that 16\% of the dependency migrations were a part of a bug fixing activity.
For example, in one of the projects, a developer created a pull request to fix a bug by migrating to an alternative package~\cite{Quixada_PR_Fix:online}.
He describes the issue as the following:
\say{isomorphic-fetch has a bug that prevents it from running in a react native environment. Since it is no longer maintained, it will never be fixed. That also means dependencies are outdated. cross-fetch is React Native compatible.}
In 8\% of the cases, we observe that the dependency migration occurs when developers add a new feature.
An example of such a case, a developer migrated to an alternative package to support the out of the box installation on more platforms~\cite{Pakers_PR_Feature:online}:
\say{And I am not arguing that it is not possible to install keystone on Windows, but this is far from a simple npm install. And as bcrypt is the problem, switching to bcryptjs would make it easier.}
In the remaining cases (6\%), we notice that the dependency migration was a part of refactoring activity.
For example, a developer refactored the code to use plain TypeScript and migrated from depending on the package \texttt{rimraf} to the package \texttt{del-cli}.

\conclusion{
    \textbf{Summary of RQ2:}
    The majority of dependency migrations (74\%) are performed to replace unmaintained dependencies, followed by compatibility issues (15\%) and performance problems (8\%). 
    Developers often migrate packages in dedicated dependency-related pull requests (69\%), but also migrate while fixing bugs (16\%), including new features (8\%), and refactoring (7\%).    
}

\section{Usefulness of the Approach}
\label{sec:alternatives:usefulness}
In this section, we evaluate the usefulness of our approach. 
We ask practitioners if they find our suggestions useful and practical for the maintenance of their software projects.
More specifically, we want to know if our approach presents new information to  practitioners: Are practitioners aware of the suggested migrations? Do practitioners believe our approach is valuable? Would practitioners act upon the suggested changes?

\subsection{Approach}

We conducted a survey to collect feedback from JavaScript project maintainers. In the following, we will present our survey design, participant recruitment process, the background of the participants, and the survey results.

\begin{table*}
    \centering
    \caption{Questions in our survey about the alternative package suggestions.}
    \label{tab:survey_questions}
    \begin{tabular}{l|p{7.3cm}|p{7.7cm}}
    \toprule
    \multirow{2}{*}{\textbf{Category}} & \multirow{2}{*}{\textbf{Question}}                                                                            & \multirow{2}{*}{\textbf{Accepted Answers}}                                                                                                                                                                     \\
                                       &                                                                                                               &                                                                                                                                                                                                                \\\midrule
    \multirow{10}{*}{Background}       & How would you best describe yourself?                                                                         & \textit{Single selection options:} Full-time, Part-time, Free-lancer, or Other.                                                                                                                                \\\cmidrule{2-3}
                                       & For how long you have been developing software?                                                               & \multirow{3}{7.7cm}{\textit{Single selection options:} Less than 1 year, 1 to 3 years, 4 to 5 years, or More  than 5 years.}                                                                                     \\\cmidrule{2-2}
                                       & How many years of JavaScript development experience do you have?                                              &                                                                                                                                                                                                                \\\cmidrule{2-2}
                                       & How many years of experience do you have using the Node Package Manager (\npm)?                               &                                                                                                                                                                                                                \\\cmidrule{2-3}
                                       & How often do you search for \npm package alternatives?                                                        & \textit{Single selection options:} Never, Rarely (once a year), Sometimes (once a month), Often (once a week), or Very often (everyday).                                               \\\midrule
    \multirow{4}{*}{Awareness}         & Are you aware of the alternative package mentioned in the email?                                              & \multirow{2}{7.7cm}{\textit{Single selection options:} Yes or No.}                                                                                                                                                 \\\cmidrule{2-2}
                                       & Are you aware of the JavaScript community's migration trend mentioned in the   email?                         &                                                                                                                                                                                                                \\\midrule
    \multirow{6}{*}{Usefulness}        & Do you think that a tool that helps generate potential alternative packages would be useful?                  & \textit{Likert-scale:} ranges from 1~$=$~Not useful, to 5~$=$~Extremely useful.                                                                                                                    \\\cmidrule{2-3}
                                       & Do you believe that providing an example of Pull Requests of migrations from other projects would be helpful? & \textit{Multiple selection options:} Help in estimating the dependency migration efforts, Help in justifying the dependency migration, Help in understanding the required API changes, Not helpful, and Other. \\\midrule
    \multirow{6}{*}{Future Actions}    & In your future new projects, will you use the alternative package?                                            & \textit{Single selection options:} Yes or No.                                                                                                                                               \\\cmidrule{2-3}
                                       & If the previous answer was "No", why not?                                                                     & \textit{Free text}                                                                                                                                                                                             \\\cmidrule{2-3}
                                       & In your current projects, will   you advise to migrate to the alternative package?                            & \textit{Likert-scale:} ranges from 1~$=$~Keep the current package, to 5~$=$~Strongly advise migrating.                                                                                                                    \\\bottomrule
\end{tabular}

\end{table*}

\noindent
\textbf{Survey Design}
We design a survey to evaluate the usefulness of the suggested dependency migration to JavaScript developers that use the packages \decline in their projects.
That is, we target developers that have used the packages our approach recommends replacing.
 \Cref{tab:survey_questions} shows the questions we ask along with the types of accepted answers for each question.
 Our survey contains three main parts, we ask JavaScript practitioners:

\begin{enumerate}
    \item Their \textbf{software development background}, to ensure that our survey participants have sufficient experience using \npm packages in software development and maintenance.

    \item Their \textbf{awareness of suggested alternatives}, to assess whether our tool can be used to inform developers of migration patterns in the ecosystem. 

    \item Their \textbf{perceptions of our suggested dependency migrations}, to assess the degree of usefulness of our approach in a real scenario. We ask participants two groups of questions: 1) how useful they found our suggested dependency migrations, and 2) their willingness to take actions related to our suggestions. In this section, we also included open-ended questions to give our survey participants the flexibility to express their opinion and experience, as recommended in survey design guidelines~\cite{dillman2011mail}.
\end{enumerate}

\noindent
\textbf{Participant Recruitment}
To select our survey participants, we reach out to experienced developers who have adopted the packages that we suggest being replaced with alternatives.
We retrieve a list of JavaScript projects hosted on GitHub that have at least 100 stars, as commonly done in the related literature~\cite{Abdalkareem_EMSE2020,Golubev_MSR2018,Borges_ICSME2016}, to mitigate the chances of including too many immature and personal projects in our survey.
We retrieve a list of 35,719 projects using the GitHub API~\cite{GitHubAPI}

Next, we use the GitHub raw content API~\cite{GitHubAPI}
to retrieve the list of dependencies from the \texttt{package.json} file of each project.
If a project has any dependencies that our approach suggests being replaced, we clone the project's repository to be analyzed.
To select participants with experience in the target dependencies, we target the developers who introduced these dependencies in their projects.
Thus, we use \texttt{git}
to retrieve the change history of the \texttt{package.json} file.
On each commit that modifies the \texttt{package.json} file, we detect the dependencies that were added, by comparing commit diffs. 
If the commit is adding a dependency that our approach suggests being replaced, we retrieve the author contacts from the commit metadata.

To prevent sending teams multiple survey invitations and to diversify our participants, we only select one developer per GitHub organization.
Based on these steps, we identify 4,696 unique JavaScript developers.
Then, we randomly selected 1,000 developers to participate in our survey.
Finally, we send email invitations of our survey to 1,000 JavaScript developers and successfully reached to 886 developers (some emails were not delivered, e.g., email address not found). 
We received 52 responses for our survey in the first two weeks, leading to a 6\% response rate, comparable to the response rate reported in other software engineering surveys~\cite{Buse_ICSE2012,SmithCHASE2013}.

\begin{table*}
    \centering
    \caption{Participants' position and experience in software development, JavaScript development, and using \npm.}
    \label{tab:survey_background}
    
\begin{tabular}{lrl|lrl|lrl|lrl}
	\toprule
	\textbf{\begin{tabular}[|c]{@{}l@{}}Developers'\\ Position\end{tabular}} & \multicolumn{2}{c|}{\textbf{Occurrences}} & \textbf{\begin{tabular}[c]{@{}l@{}}Development\\ Experience\end{tabular}} & \multicolumn{2}{c|}{\textbf{Occurrences}} & \textbf{\begin{tabular}[c]{@{}l@{}}Experience\\ in JavaScript\end{tabular}} & \multicolumn{2}{c|}{\textbf{Occurrences}} & \textbf{\begin{tabular}[c]{@{}l@{}}Experience in\\ Using \npm\end{tabular}} & \multicolumn{2}{c}{\textbf{Occurrences}}                                                        \\ \midrule
	Full-time                          & 42                                        & \sbar{42}{52}                      & 1 - 3                                     & 0                                  & \sbar{0}{52}                              & 1 - 3                              & 3                                        & \sbar{3}{52}  & 1 - 3           & 3  & \sbar{3}{52}  \\
	Part-time                          & 1                                         & \sbar{1}{52}                       & 4 - 5                                     & 5                                  & \sbar{5}{52}                              & 4 - 5                              & 8                                        & \sbar{8}{52}  & 4 - 5           & 9  & \sbar{9}{52}  \\
	Freelancer                         & 9                                         & \sbar{9}{52}                       & ~\textgreater~5                           & 47                                 & \sbar{47}{52}                             & ~\textgreater~5                    & 41                                       & \sbar{41}{52} & ~\textgreater~5 & 40 & \sbar{40}{52} \\ \bottomrule
\end{tabular}

\end{table*}

\noindent
\textbf{Survey Participants.}
\Cref{tab:survey_background} shows the background of our survey participants, including their position, their experience in software development, JavaScript development, and the use of the \npm packages.
Overall, the majority of participants work full-time (42 out of 52), and have at least 5 years of experience in software development in general (47 out of 52), JavaScript development (41 out of 52), and using \npm (40 out of 52).

\begin{figure}
    \centering
    \includegraphics[width=.85\linewidth]{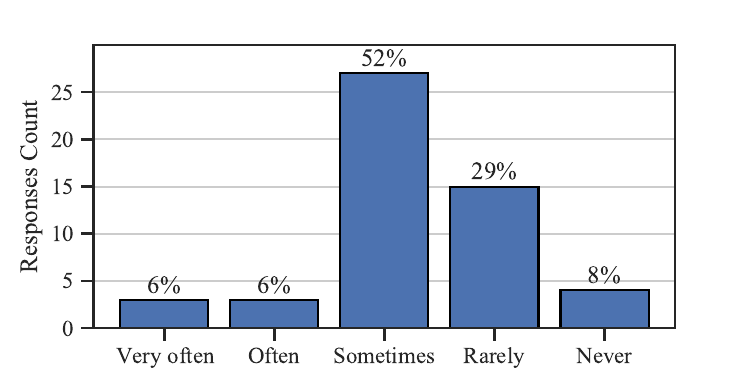}
    \fullcaption[Survey responses to how often our would participants search for package alternatives.]{The question has the following answers: never, rarely (once a year), sometimes (once a month), often (once a week), very often (everyday).}
    \label{fig:search_for_alternatives}
\end{figure}

We also ask participants how often they search for alternative \npm packages, to evaluate their interest and experience in finding better \npm packages for their projects. 
\Cref{fig:search_for_alternatives} show that out of the 52 participants, 92\% of them do search for alternatives \npm packages, consolidating this task as a common maintenance task in the maintenance of software projects. 
The frequency in which developers look for other packages varied from once a month (52\%), once a year (29\%), once a week (6\%), and every day (6\%).
Interestingly, only 8\% of our survey participants report that they never search for alternative packages.

\subsection{Results}
We measure our approach usefulness along three dimensions: 1)~developers awareness of the generated suggestions, 2)~developers perceptions of our suggestions, and 3)~developers willingness to take future actions based on our suggestions.

\begin{figure}
    \centering
    \includegraphics[width=\linewidth]{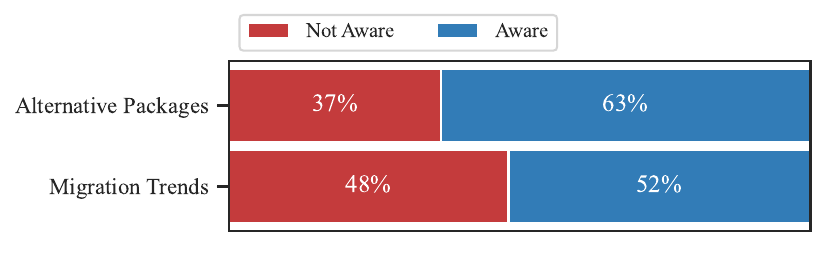}
    \caption{The awareness of participants about the suggested alternative packages and the community's migration trends.}
    \label{fig:awareness}
\end{figure}

\mainpoint{Developer Awareness}
We assess the awareness about the generated suggestions by asking the participants 1) if they know the alternative package and 2) whether they are aware of the migration trend in the JavaScript community toward the alternative packages.
Based on our survey responses, our approach was able to inform participants about the alternative packages and the JavaScript community's migration trend.
Specifically, \Cref{fig:awareness} shows that 37\% of the participants are not aware of suggested alternative packages, and 48\% of the participants are not aware of the dependency migration trend in the JavaScript community.
Given our suggestions are based on migrations that have been performed many times in the \npm ecosystem, we found it surprising that 37\% of participants had not heard of the alternative package before. 
To put things into perspective, all of our survey participants are familiar with the original package and 92\% have reported to regularly looking in the ecosystem for alternative packages.
This indicates that even experienced developers are frequently unaware of the ecosystem' trends and need tools to be better informed about their community.

\begin{figure}
    \centering
    \includegraphics[width=.8\linewidth]{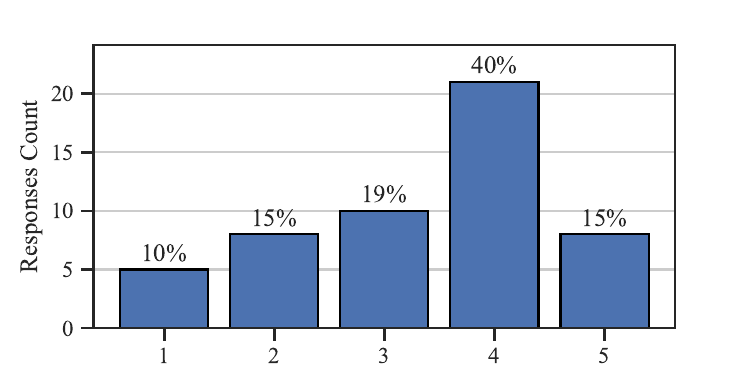}
    \fullcaption[Survey responses to the usefulness of having a tool that generate alternative suggestions using our approach.]{The usefulness rated on a 5-points Likert-scale
        ranges from 1~$=$~Not useful, to 5~$=$~Extremely useful.}
    \label{fig:usefulness_tool}
\end{figure}

\mainpoint{Developer Perceptions}
To understand the perception of the project maintainers of our approach's results, we ask participants if they think that a tool that generates potential alternative packages would be useful.
As \Cref{fig:usefulness_tool} shows, most participants believe that the suggestions generated by our approach are useful and support the idea of having a tool to generate such suggestions.
On a 5-points Likert scale, the support of such a tool has a median~$=$~4 and mean~$=$~3.37, where only 10\% of the participants claimed it is not useful for them, and 15\% indicate that it is extremely useful.

\begin{table}
    \centering
    \caption{Participants' responses on how helpful are the examples of dependency migrations from other projects?}
    \label{tab:pr_usefulness}
    \begin{tabular}{l|rl}
    \toprule
    \textbf{Helpfulness}                        & \multicolumn{2}{c}{\textbf{Frequency}}                 \\
    \midrule
    Understanding the required API changes      & 79\%                                   & \sbar{41}{52} \\
    Estimating the dependency migration efforts & 75\%                                   & \sbar{39}{52} \\
    Justifying the dependency migration         & 52\%                                   & \sbar{27}{52} \\
    Other                                       & 6\%                                    & \sbar{3}{52}  \\   \midrule
    Not helpful                                 & 11\%                                   & \sbar{6}{52}  \\
    \bottomrule
\end{tabular}

\end{table}

In the survey invitation, we provide the participants with pull request examples of dependency migrations from other projects.
We ask the participants if they found the pull request examples helpful.
As shown in \Cref{tab:pr_usefulness}, the majority of the participants (88\%) believe that the provided examples of dependency migrations from other projects are helpful.
Specifically, 79\% of the participants find that the provided examples help understand the differences and the required changes in the API usage between the alternative package and the current package.
Also, 75\% of the participants indicate that the migration examples from other projects can help in estimating the efforts needed to migrate to the alternative package in their projects.
Interestingly, 52\% of the participants mentioned that the explanation in the provided examples helps justify the dependency migration within their teams.
Finally, only 12\% of the participants did not find the examples helpful for their project. 
Participant P3 expressed a disagreement with the justification given in the pull request example.
The participants believe that the alternative package has more dependencies, which is the opposite to what was explained in the pull request example, \say{Just reading briefly, but yargs has way more dependencies than commander, contrary on what is reported in the PR}.
Another participant mentioned that a pull request example would only be useful to them if it illustrates solving a security issue (P50:~\say{I would never bother migrating unless there were a severe and applicable security concern~...}).

\begin{figure}
    \centering
    \includegraphics[width=.8\linewidth]{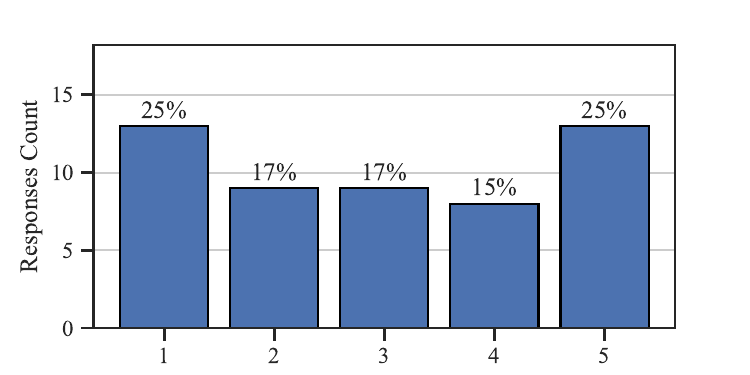}
    \fullcaption[Survey responses on the support of migrating their current projects to use the alternative packages.]{The support rated on a 5-points Likert-scale ranges from 1~$=$~Keep the current package, to 5~$=$~Strongly advise migrating.}
    \label{fig:action_current}
\end{figure}

\mainpoint{Future Actions}
Whether developers would act upon our suggestions is a strong indication of our approach's usefulness.
We ask the participants 1) if they plan to use the alternative packages in their future projects and 2) how likely are practitioners to migrate to alternative packages in their current projects. 
The responses from our survey shows that 67\% of the participants will use the alternative packages in their future projects.
However, participants are split between supporting the dependency migrations and keeping current packages in their current project (see \Cref{fig:action_current}).
On a 5-point Likert scale, participants rate their support to migrate to the alternative package in their current projects in median~$=$~2.

To understand why participants opt for retaining current packages, even if the package has been declining in the community, we analyze the free-text justifications from our survey participants on why they will not use the alternative package.
In total, we manually examined 17 responses (out of the 52 responses) from the participants who indicated that they would not use the alternative package.
Based on this analysis, we observe different reasons why they will not use the alternative package.
Six participants mentioned that they would not use an alternative package as long as the current one is working.
As one respondent P2 summarizes, \say{If it ain't broke, don't fix.}
While migrating to a dependency with more functionalities can be helpful, the benefits of migrating will only be achieved if the project needs the extra functionality.
One of our participants (P47) expressed this concern and said \say{The alternative packages bundles many functions, I need just the one that the old package uses.}
In contrast, P3 prefers not to use a package that has more dependencies, and state \say{Because I value packages with little to none dependencies, one reason (of many) is the Dependabot's alerts hell, which is a huge waste of time, more often than not.}
Other participants indicated having less priority to migrating packages that perform a minor functionality or packages that they do not use in the production.
For example, P9 said \say{It's just a small development time dependency used during build...as long it works - it works.}
Interestingly, some participants will not use the alternative packages because their current projects have low priority. For example, P13 said \say{The project is mostly deprecated and will be moving to a new golang based system.}

\diego{@Suhaib: This summary needs to be improved}
\conclusion{
    \textbf{Summary of RQ3:}
    Our approach was deemed very useful for practitioners (median of 4 in 5-point Likert scale) as it showcases new alternative packages, informs about migration trends, and help estimate migration efforts.    
    Out of our 52 participants, 67\% said they consider using suggested alternatives in their future projects but were more conservative when migrating current projects due to potential maintenance risks. 
    
}

\section{Discussion}
\label{sec:implications}
In this section, we discuss our results and how our approach may help improve software maintenance activities. 

\noindent
\textbf{The need for tools that increase community awareness.} 
Our RQ3 results strongly suggested that even experienced JavaScript developers, that frequently search for new and better alternatives for their project dependencies, cannot keep track of ecosystem evolution. 
Out of the 52 respondents, almost half (48\%) were not aware of migration trends, and 37\% did not even know of the better alternative package. 
Given that, migrations are primarily motivated by better software maintainability (RQ2), there is a dire need (and opportunity) to harness complex community patterns to inform practitioners on potential maintenance problems.
By recommending better alternative packages, our tool can effectively inform developers that the packages they use have become poorly maintained. 
Poorly maintained dependencies are a major maintenance risk for software projects, as they are unlikely to respond to bug fixing request or reported vulnerabilities~\cite{Zajdel:OSS,Alfadel:Dependabot}.

\noindent
\textbf{Our approach is scalable and would fit well in open source workflow.}
While our approach entails mining an entire ecosystem to identify migration patterns, the computation power needed to identify such patterns is modest. 
Analyzing the history of dependency changes in the entire \npm ecosystem took approximately 24 hours, and new suggestions can be incrementally analyzed in minutes. 
Our approach can be computed centrally on a server, and be provided as a service to subscribers - similarly to Dependabot~\cite{Alfadel:Dependabot}, informing maintainers when a new migration suggestion is identified.
Given our RQ2 showed that, migrations are frequently performed in dedicated tasks, developers can validate suggestions - discard the ones not relevant for their project - and perform the desired migrations in a batch, before project releases.

\noindent
\textbf{Thresholds and tool configuration.}
Identifying useful patterns using the wisdom of the crowd requires navigating the delicate trade-off between the sensitivity and the quality of the migration suggestion.
We opted to identify fewer but better quality suggestions by choosing the criteria shown in Section~\ref{sub:criteria}.
A much larger number of suggestions would be extracted, for example, if we relaxed the need to have the migration be performed recently, however, at the cost of recommending outdated migration suggestions. 
While a more in-depth analysis may identify an even better set of criteria, our criteria is a good starting point for further exploration.

\section{Related Work}
\label{sec:alternatives:related_work}

Several studies proposed approaches to recommend packages to developers.
\citet{Thung_WCRE2013} proposed an approach to recommend packages for projects based on their current dependencies, using association rule mining and collaborative filtering.
Other studies targeted the same problem by using different approaches, such as multi-objective optimization~\cite{Ouni_IST2017}, and pattern mining and hierarchical clustering~\cite{Saied_JSS2018}.
Recently, \citet{Nguyen_JSS2020} proposed a more efficient approach as it generates recommendations in a comparably less historical data. 
The main goal of these approaches is to tap in the missed opportunities of using available packages, based on the project's package dependencies and characteristics.
However, our goal in this study is to recommend migration opportunities of better alternatives than packages already in use.

\citet{Chen_SANER2016} proposed an approach for mining Stack Overflow tags to find semantically similar packages.
Even though this approach can return a set of similar packages, it has no evidence of the feasibility of migrations between the alternatives~\cite{He_SANER2021}.
Thus, researchers proposed mining historical migrations from existing software repositories, which rely on the crowd's wisdom in performing migrations to alternative packages~\cite{Teyton_WCRE2012,Teyton_SMR2014,Alrubaye_ICSE219}.
However, their approaches suffer from either low recall~\cite{Teyton_WCRE2012,Alrubaye_ICSE219}, or low precision \cite{Teyton_WCRE2012,Teyton_SMR2014}.
\citet{He_SANER2021} improved the performance by utilizing multiple metrics to capture different dimensions of evidence from development histories when recommending dependency migrations extracted from other software repositories.
Since this approach relies on analyzing the commits and their messages to extract migration patterns, it is sensitive to how developers divide their changes across commits and the clarity of commit messages.
In contrast, our approach extracts dependency migrations based on versions without considering the individual commits, which overcomes the previous limitation.
Also, our approach targets migration suggestions for packages \decline only,  avoiding the problem of undesired suggestions~\cite{Erlenhov_FSE2020}.

Researchers empirically investigated dependency migrations.
\citet{Kabinna_MSR2017} highlight the challenges in migrating to new logging packages.
\citet{Alrubaye_ICSR2020} analyzed several code quality metrics before and after applying dependency migrations.
Others studied the role of common metrics in developer selection of packages~\cite{Mora_PROMISE2018, Mujahid_JSS2023}.
\citet{He_SANER2021} proposed new four metrics (i.e., Rule Support, Message Support, Distance Support, and API Support) to rank the migration suggestions.
They all use project level metrics in their approaches, while we are the first to use the centrality in the context of dependency migrations (i.e., an ecosystem level metric), which emphasizes the community interest in performing the dependency migrations.

\section{Threats to Validity}
\label{sec:alternatives:threats_to_validity}

\noindent
\textbf{Threats to Internal Validity.}
Threats to internal validity are related to experimenter bias and errors.
A limitation of our approach is that it only considers dependencies between packages in the \npm registry.
This limitation will affect the centrality and migration patterns of packages that are not meant to be used by other packages, but other JavaScript applications, i.e., top-level packages.
However, previous work has shown that using the \npm registry as the sole source of changes in the dependency graph can serve as a proxy for the overall~\cite{Mujahid_TEM2021,Cogo_TSE2021,Cogo_TSE2019}.
Future work should investigate how to incorporate JavaScript applications on the generated suggestions.
Another threat concerns the process we used to filter dependency change events to our approach.
To reduce noise, we opted to remove dependency change events from massive or imbalanced number dependency changes, as explained in \Cref{sub:detect_dependency_events}.
This may affect our recommended patterns, as they will more likely based in projects that perform small dependency changes over time.
We mitigate this effect by only considering migration patterns that reoccur across many projects.
Finally, our approach may contain bugs that may have affected our results.
We made our scripts and dataset publicly available to be fully transparent and have the community help verify (and improve) our approach~\cite{suhaib_mujahid_2021_5548231}.

\noindent
\textbf{Threats to External Validity.}
Our evaluation focused entirely on the \npm ecosystem, which has very particular characteristics: a centralized package registry, hundreds of thousands of software packages, and a very active and popular programming language.
Also, packages in the \npm ecosystem are relatively small compared to modules and software components in other ecosystems and programming languages~\cite{Decan_EMSE2019}, which could lead to different dynamics than other ecosystems, significantly affecting the dependency migration patterns. 
Future work needs to investigate if a similar approach can effectively find dependency migration suggestions in other ecosystems such as PyPI and Maven.

\section{Conclusion}
\label{sec:alternatives:conclusion}

This paper presents an approach to extract dependency migration trends in the software ecosystem and suggests alternatives for packages that are \decline.
We evaluate our approach in \npm, one of the largest and most popular software ecosystems.
Our evaluation showed that our approach was accurate at suggesting alternative packages (RQ1), recommended migrations that were primarily motivated by maintenance issues (RQ2), and found that developers support having a tool that utilizes our approach to suggest alternative packages (RQ3).
Future work could explore avenues to improve and better operationalize our proposed approach. 
While we generate suggestions for one-to-one dependency migrations, it is valuable for future work to support one-to-many and many-to-many dependency migrations, where one or more packages are replacing one or more packages.
Another exciting follow-up work is to propose an automated approach that uses the dependency migration examples to perform the suggested dependency migrations, refactoring the code, and ensuring the semantics are preserved during the migration.

\noindent
\textbf{Acknowledgement}.
The involvement of human participants in this study was approved by Concordia University’s Research Ethics Board (protocol form number: 3000472)

\bibliographystyle{IEEEtranN}
\balance
{\small
\bibliography{main.bib}}
\onecolumn
\listoftodos{}

\end{document}